\documentstyle[epsf,aps,prb]{revtex}
\def\eg{for example}
\def\b{\beta}
\def\pd#1#2{{\partial #1\over\partial #2}} 
\def\p2d#1#2{{\partial^2 #1\over\partial #2^2}} 
\def\erf{{\rm erf}}\def\erfc{{\rm erfc}}
\def\av#1{\langle #1\rangle} 
\def\CR{r} 
\def\pnd#1#2#3{{\partial^{#3} #1\over\partial #2^{#3}}} 
\def\CC{F} 
\def\CD{\hbox{{$\cal D$}}}
\def\td#1#2{{d #1\over d #2}} 
\def\t2d#1#2{{d^2 #1\over d #2^2}} 

\def\ap #1 #2 #3 {{\sl Adv.\ Phys.} {\bf #1}, #2 (#3)}
\def\jpa #1 #2 #3 {{\sl J. Phys.\ A} {\bf #1}, #2 (#3)}
\def\ajp #1 #2 #3 {{\sl Am.\ J.\ Phys.} {\bf #1}, #2 (#3)}
\begin{document}
\title{Capture of the Lamb: Diffusing Predators Seeking a
Diffusing Prey}
\medskip
\author{S. Redner and P.~L.~Krapivsky}
\medskip
\address{Center for BioDynamics, Center for Polymer Studies,
and Department of Physics,\\
Boston University, Boston, MA 02215}

\maketitle

\begin{abstract}
 
 We study the capture of a diffusing ``lamb'' by diffusing ``lions'' in one
 dimension. The capture dynamics is exactly soluble by probabilistic
 techniques when the number of lions is very small, and is tractable by
 extreme statistics considerations when the number of lions is very large.
 However, the exact solution for the general case of three or more lions is
 still not known.

\end{abstract}

\medskip \noindent {\bf I. INTRODUCTION}

\begin{figure}
\epsfxsize=50mm\epsfysize=50mm\centerline{\epsfbox{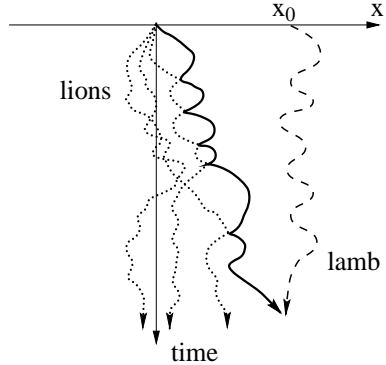}}
\vskip 0.1in
\caption{Space-time evolution in one dimension of $N=4$ diffusing
lions (dotted lines) which all start at $x=0$ and a single
diffusing lamb (dashed) which starts at $x=x_0$. The trajectory
of the closest (``last'') lion, whose individual identity may
change with time, is indicated by the heavy solid path.
\label{space-time}}
\end{figure}

What is the survival probability of a diffusing lamb which is hunted by $N$
hungry lions? Although this capture process is appealingly simple to define
(see Fig.~\ref{space-time}), its long-time behavior\cite{bg,kesten,kr} poses
a theoretical challenge because of the delicate interplay between the
positions of the lamb and the closest lion. This model also illustrates a
general feature of nonequilibrium statistical mechanics: life is richer in
low dimensions. For spatial dimension $d>2$, it is known that the capture
process is ``unsuccessful'' (in the terminology of Ref.~1), as there is a
nonzero probability for the lamb to survive to infinite time for any initial
spatial distribution of the lions. This result is a consequence of the
{\em transience} of diffusion for $d>2$,\cite{feller,weiss} which
means that two nearby diffusing particles in an unbounded $d>2$
domain may never meet. For
$d=2$, capture is ``successful'', as the lamb dies with certainty. However,
diffusing lions in $d=2$ are such poor predators that
the average lifetime of the lamb is infinite! Also, the lions are
essentially independent,\cite{bg} so that the survival probability
of a lamb in the presence of $N$ lions in two dimensions is
$S_N(t)\propto S_1(t)^N$, where $S_1(t)$, the survival probability
of a lamb in the presence of a single lion, decays as\cite{weiss}
$(\ln t)^{-1}$.

Lions are more efficient predators in $d=1$ because of the {\em
recurrence} of diffusion,\cite{feller,weiss} which means that two
diffusing particles are certain to meet eventually. The $d=1$ case
is also special because there are two distinct generic cases. When
the lamb is surrounded by lions, the survival probability at a
fixed time decreases rapidly with $N$ because the safe zone which
remains unvisited by lions at fixed time shrinks rapidly in
$N$. This article focuses on the more interesting situation of $N$ lions all
to one side of the lamb (Fig.~\ref{space-time}), for which the lamb survival
probability decays as a power law in time with an exponent that grows only
logarithmically in $N$.

We begin by considering a lamb and a single stationary lion in
Section~II. The survival probability of the lamb $S_1(t)$ is
closely related to the first-passage probability of one-dimensional
diffusion\cite{feller,weiss} and leads to $S_1(t)\sim t^{-1/2}$. It
is also instructive to consider general lion and lamb
diffusivities. We treat this two-particle system by mapping it onto
an effective single-particle diffusion problem in two dimensions
with an absorbing boundary to account for the death of the lamb
when it meets the lion, and then solving the two-dimensional
problem by the image method. We apply this approach in Section~III
by mapping a diffusing lamb and two diffusing lions onto a
single diffusing particle within an absorbing wedge whose opening
angle depends on the particle diffusivities,\cite{dba} and then
solving the diffusion problem in this absorbing wedge by classical
methods.

In Section IV, we study $N\gg 1$ diffusing lions.\cite{kesten,kr} An
essential feature of this system is that the motion of the closest (``last'')
lion to the lamb is biased towards the lamb, even though each lion diffuses
isotropically. The many-particle system can be recast as a two-particle
system consisting of the lamb and an absorbing boundary which, from extreme
statistics, \cite{extreme} moves to the right as $\sqrt{4D_L t\,\ln N}$,
where $D_L$ is the lion diffusivity. Because this time dependence matches
that of the lamb's diffusion, the survival probability depends intimately on
these two motions,\cite{movingm,movingp,movingkr} with
the result that $S_N(t)\sim t^{-\b_N}$ and $\b_N\propto\ln N$. The
logarithmic dependence of $\b_N$ on
$N$ reflects the fact that each additional lion poses a progressively smaller
marginal peril to the lamb --- it matters little whether the lamb is hunted
by 99 or 100 lions. Amusingly, the value of $\beta_N$ implies an
infinite lamb lifetime for $N\leq 3$ and a finite lifetime
otherwise. In the terminology of Ref.~1, the capture process
changes from ``successful'' to ``complete'' when
$N\geq 4$. We close with some suggestions for additional research on this
topic.

\medskip\noindent {\bf II. SURVIVAL IN THE PRESENCE OF ONE LION}

\smallskip \noindent {\bf A. Stationary Lion and Diffusing
Lamb}

We begin by treating a lamb which starts at $x_0>0$ and a
stationary lion at $x=0$. In the continuum limit, the probability
density $p(x,t)$ that the lamb is at any point $x>0$ at time $t$
satisfies the diffusion equation
\begin{equation}
\label{de}
\pd{p(x,t)}{t}=D_\ell\,\p2d{p(x,t)}{x},
\end{equation}
where $D_\ell$ is the diffusivity (or diffusion coefficient). The
probability density satisfies the boundary condition
$p(x=0,t)=0$ to account for the death of the lamb if it reaches
the lion at
$x=0$, and the initial condition
$p(x,t=0)=\delta(x-x_0)$. Equation~(\ref{de}) may be easily solved
by the familiar image method.\cite{weiss} For $x>0$, $p(x,t)$ is
the superposition of a Gaussian centered at $x_0$ and an ``image''
anti-Gaussian centered at $-x_0$:
\begin{equation}
\label{de-soln}
p(x,t)={1\over\sqrt{4\pi D_\ell t}}\left[e^{-(x-x_0)^2/4D_\ell t}-
 e^{-(x+x_0)^2/4D_\ell t}\right].
\end{equation}
The image contribution ensures that the boundary condition at $x=0$ is
automatically satisfied, while the full solution satisfies both the initial
condition and the diffusion equation. Thus Eq.~(\ref{de-soln}) gives the
probability density of the lamb for $x>0$ in the presence of a stationary
lion at $x=0$.

The probability that the lamb meets the lion at time $t$ equals the
diffusive flux to $x=0$ at time $t$. The flux is 
\begin{equation}
\label{fpp}
F(t) = +D_\ell\pd{p(x,t)}{x}\bigg|_{x=0}
 = {x_0\over\sqrt{4\pi D_\ell t^3}}\,\,e^{-x_0^2/4D_\ell t}. 
\end{equation}
The flux $F(t)$ is also the {\em first-passage probability} to the
origin, namely, the probability that a diffusing lamb which starts
at $x_0$ reaches
$x=0$ {\em for the first time} at time $t$. Note that in the long
time limit, defined by $D_\ell t\gg x_0^2$, the first-passage
probability reduces to $F(t)\to x_0/t^{3/2}$. This
$t^{-3/2}$ time dependence is a characteristic feature of the
first-passage probability in one dimension.\cite{weiss}

The probability that the lamb dies by time $t$ is the time
integral of $F(t)$ up to time $t$. The survival probability is
just the complementary fraction of these doomed lambs, that is,
\begin{eqnarray}
 \label{Svst}
S_1(t) &=& 1-\! \int_0^t \! F(t')\,dt', \nonumber \\
&=& 1-\! \int_0^t \! {x_0\over\sqrt{4\pi D_\ell t'^3}} \,\,
e^{-x_0^2/4D_\ell t'}\, dt'.
\end{eqnarray}
The integral in Eq.~(\ref{Svst}) can be reduced to a standard form
by the substitution
$u=x_0/\sqrt{4D_\ell t'}$ to give
\begin{equation}
 \label{Svst-soln}
 S_1(t) = \erf\biggl({x_0\over\sqrt{4D_\ell t}}\biggr)
 \sim {x_0\over\sqrt{\pi D_\ell t}}
\quad {\rm as~} t\to\infty,
\end{equation}
where $\erf(z)=(2/\sqrt{\pi}) \! \int_0^z e^{-u^2}\,du$ is the error
function.\cite{as} The same expression for $S_1(t)$ can be obtained by
integrating the spatial probability distribution in Eq.~(\ref{de-soln}) over
all $x>0$.

An amusing feature of the $t^{-1/2}$ decay of the lamb survival
probability is that although the lamb dies with certainty, its
average lifetime, defined as
$\av{t}=\!\int_0^\infty t\, F(t)\,dt=\int_0^\infty S(t)\,dt \approx
\int^\infty t^{-1/2}\, dt$, is infinite. This infinite lifetime
arises because the small fraction of lambs which survive tend to
move relatively far away from the lion. More precisely, the
superposition of the Gaussian and anti-Gaussian in
Eq.~(\ref{de-soln}) leads to a lamb probability distribution which
is peaked at a distance $(D_\ell t)^{1/2}$ from the origin, while
its spatial integral decays as $(D_\ell t)^{-1/2}$.

\smallskip \noindent{\bf B. Both Species Diffusing}

What is the survival probability of the lamb when the lion also diffuses? In
the rest frame of the lamb, the lion now moves if either a lion or a
lamb hopping event occurs, and their separation diffuses with diffusivity
equal to $D_\ell+D_L$ (see \eg, Ref.~\ref{bib:weiss}), where $D_L$
is the lion diffusivity. From the discussion of Section~IIA, the
lamb survival probability has the asymptotic time
dependence
$S_1(t)\sim x_0/\sqrt{\pi(D_\ell+D_L) t}$.

\begin{figure}
\epsfxsize=54mm\epsfysize=40mm \centerline{\epsfbox{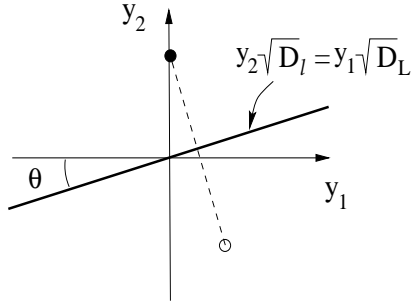}}
\vskip 0.1in
\caption{Mapping of the lion and lamb coordinates in one dimension to the planar 
 coordinates $y_1=x_L/\sqrt{D_L}$ and $y_2= x_\ell/\sqrt{D_\ell}$. The
 initial $y$-coordinates of the lion-lamb pair, $(0,\sqrt{D_\ell})$, and its
 image are indicated by the solid and open circles, respectively. Survival
 of the lamb, $y_1\sqrt{D_L}<y_2\sqrt{D_\ell}$, translates to the diffusing
 particle in the plane remaining above and to the left of the absorbing line
 $y_1\sqrt{D_L}=y_2\sqrt{D_\ell}$.
\label{plane}}
\end{figure}

It is also instructive to determine the spatial probability distribution of
the lamb. This distribution may be found conveniently by mapping
the two-particle interacting system of lion at $x_L$ and lamb at
$x_\ell$ in one dimension to an effective single-particle system in
two dimensions and then applying the image method to solve the
latter (see Fig.~\ref{plane}). To construct this mapping, we
introduce the scaled coordinates
$y_1=x_L/\sqrt{D_L}$ and $y_2=x_\ell/\sqrt{D_\ell}$ to render the
two-dimensional diffusive trajectory $(y_1,y_2)$ isotropic. The
probability density in the plane, $p(y_1,y_2,t)$, must satisfy an
absorbing boundary condition when $y_2\sqrt{D_\ell}=y_1\sqrt{D_L}$,
corresponding to the death of the lamb when it meets the lion. For
simplicity and without loss of generality, we assume that the lion
and lamb are initially at $x_L(0)=0$ and $x_\ell(0)=1$
respectively, that is,
$y_1(0)=0$ and $y_2(0)=\sqrt{D_\ell}$. The probability density is therefore
the sum of a Gaussian centered at $(y_1(0),y_2(0))=(0,\sqrt{D_\ell})$ and an
anti-Gaussian image. From the orientation of the absorbing boundary
(Fig.~\ref{plane}), this image is centered at $(\sqrt{D_\ell}\sin
2\theta,-\sqrt{D_\ell}\cos 2\theta)$, where $\theta =
\tan^{-1}\! \sqrt{D_L/D_\ell}$.

\begin{figure}
\epsfxsize=63mm\epsfysize=45mm\hskip 1.5in
\centerline{\epsfbox{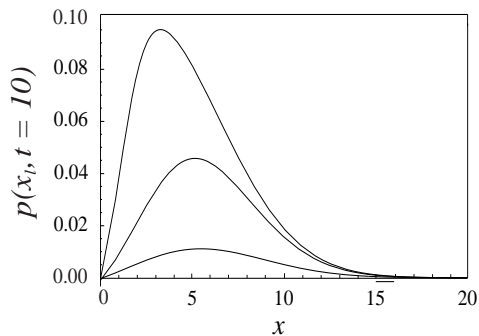}}
\vskip 0.0in
\caption{Probability distribution of the lamb in one dimension at
time $t=10$ (Eq.~(\ref{err-fn})) when the lion and lamb are
initially at
$x_L=0$ and $x_\ell=1$, respectively. The cases shown are 
$r=D_\ell/D_L=0.1$, 1, and 10 (bottom to top).}
\label{lamb-pdf}
\end{figure}

{}From this image representation, the probability density in two
dimensions is
\begin{equation}
\label{planar}
p(y_1,y_2,t) = {1\over{4\pi t}} \left
[e^{-[y_1^2+(y_2-\sqrt{D_\ell})^2]/4t} -e^{-[(y_1-\sqrt{D_\ell}\sin
2\theta)^2+ (y_2+\sqrt{D_\ell}\cos 2\theta)^2]/4t}\right ].
\end{equation}
The probability density for the lamb to be at $y_2$ is the integral
of the two-dimensional density over the accessible range of the lion
coordinate $y_1$:
\begin{equation}
\label{1d}
p(y_2,t) = \!\int_{-\infty}^{y_2\cot\theta} \!
p(y_1,y_2,t)\, dy_1.
\end{equation}
If we substitute the result (\ref{planar}) for $p(y_1,y_2,t)$, the
integral in Eq.~(\ref{1d}) can be expressed in
terms of the error function. We then transform back to the original
lamb coordinate
$x_\ell=y_2\sqrt{D_\ell}$ by using $p(x_\ell,t)\,dx_\ell=p(y_2,t)\,dy_2$ to
obtain
\begin{equation}
\label{err-fn}
p(x_\ell,t) = {1\over\sqrt{16\pi D_\ell t}}\biggl [
e^{-(x_\ell-1)^2/4D_\ell
t}\,\,\erfc\Bigl(-{x_\ell\cot\theta\over\sqrt{4D_\ell t}}\Bigr) -
e^{-(x_\ell+\cos 2\theta)^2/4D_\ell t}\,\,
\erfc\Bigl({\sin 2\theta-x_\ell\cot\theta\over\sqrt{4D_\ell
t}}\Bigr)\biggr],
\end{equation}
where $\erfc(z)=1-\erf(z)$ is the complementary error function. A plot of
$p(x_\ell,t)$ is shown in Fig.~\ref{lamb-pdf} for various values of the
diffusivity ratio $r\equiv D_\ell/D_L$. The figure shows that the survival
probability of the lamb rapidly decreases as the lion becomes more mobile.
Note that when the lion is stationary, $\theta=0$, and
Eq.~(\ref{err-fn}) reduces to Eq.~(\ref{de-soln}). 

\medskip \noindent{\bf III. TWO LIONS}

\begin{figure}
\epsfxsize=70mm\epsfysize=60mm\centerline{\epsfbox{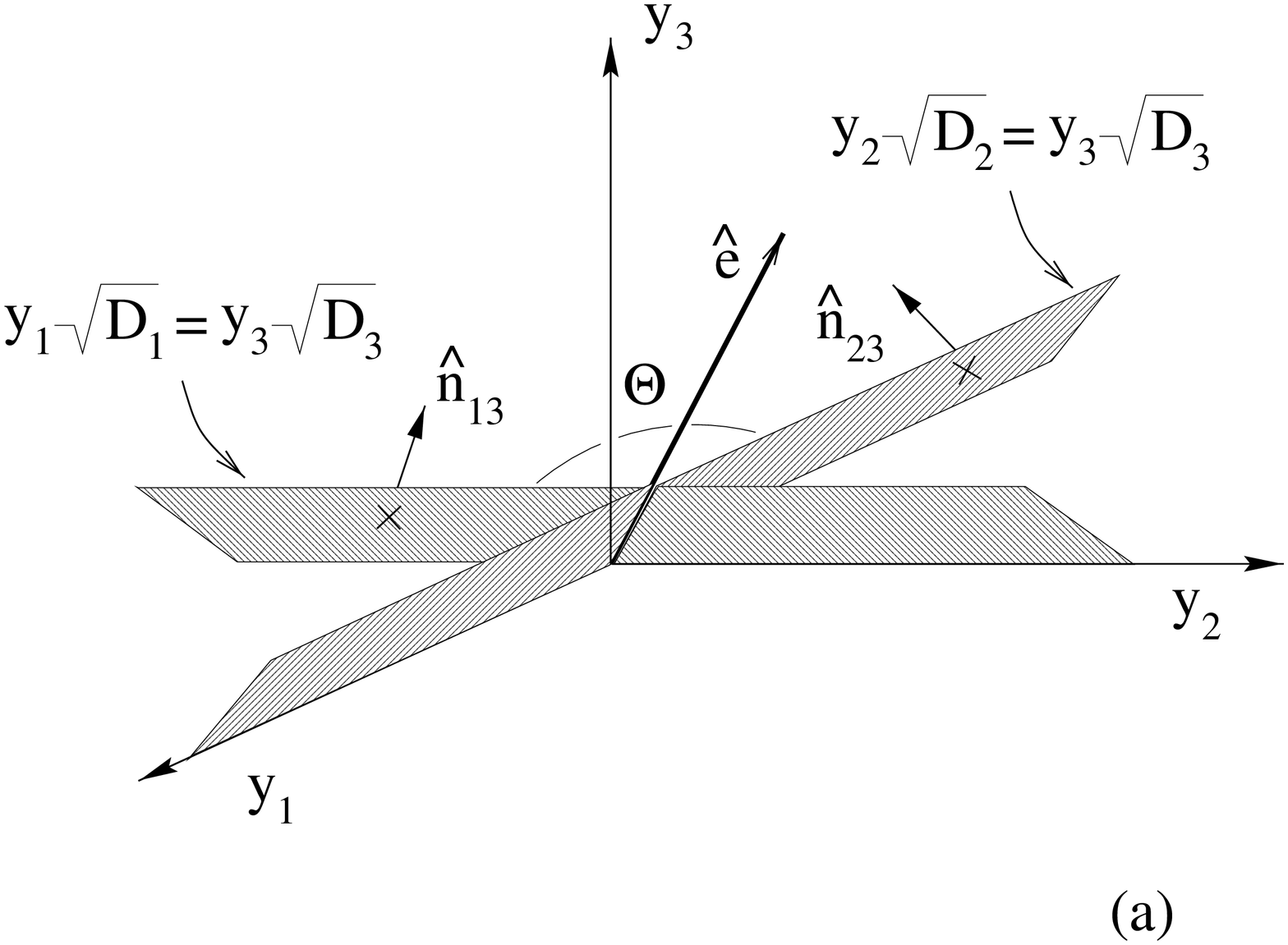}}
\vskip 0.0in
\epsfxsize=75mm\epsfysize=35mm\centerline{\epsfbox{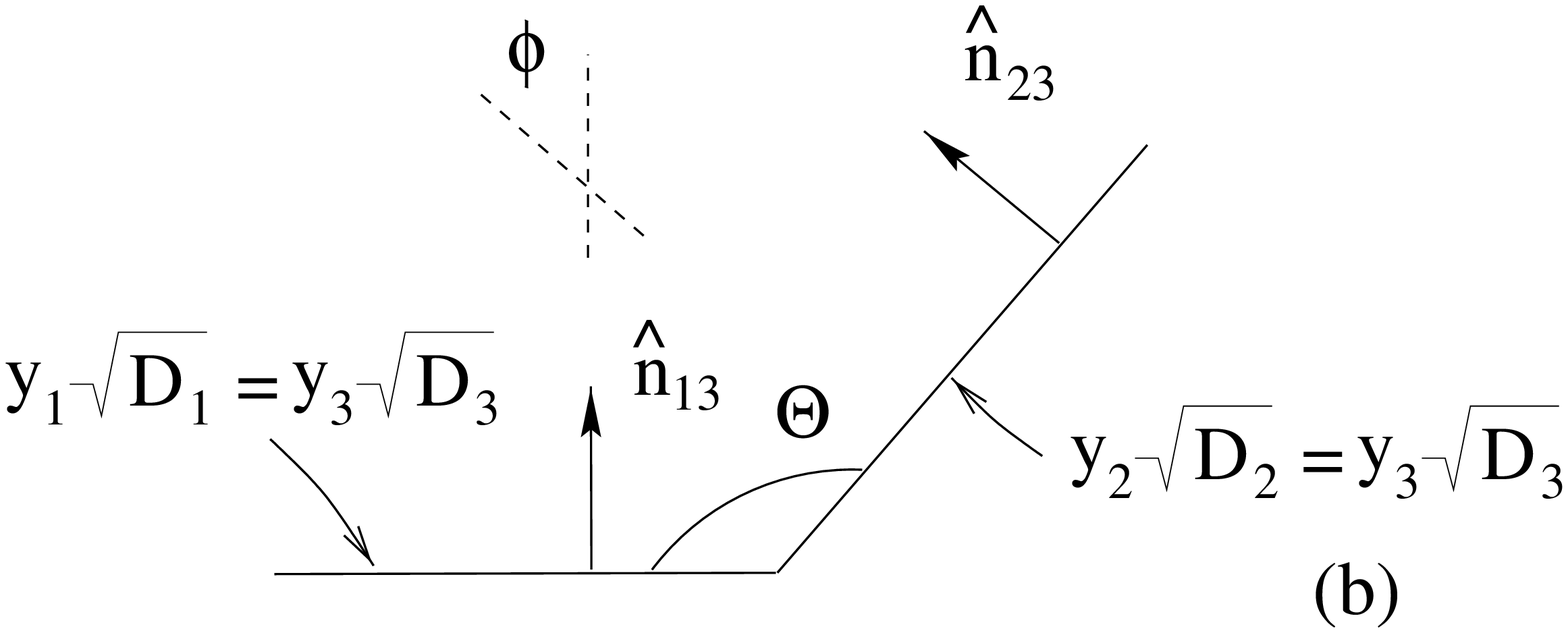}}
\vskip 0.1in
\caption{(a) Mapping between the coordinates $x_i$ of three diffusing
 particles on the line and a single isotropically diffusing particle in the
 three-dimensional space $y_i=x_i/\sqrt{D_i}$, subject to the constraints
 $y_1\sqrt{D_1}<y_3\sqrt{D_3}$ and $y_2\sqrt{D_2}<y_3\sqrt{D_3}$. The lamb
 survives if it remains within the wedge-shaped region of opening angle
 $\Theta$. (b) Projection of the wedge onto a plane perpendicular to the
 $\hat e$ axis defined by the intersection of the two planes.
\label{wedge}}
\end{figure}

To find the lamb survival probability in the presence of two diffusing lions,
we generalize the above approach to map the three-particle interacting system
in one dimension to an effective single diffusing particle in three
dimensions with boundary conditions that reflect the death of the lamb
whenever a lion is encountered.\cite{dba} Let us label the lions as particles
1 and 2, and the lamb as particle 3, with respective positions $x_1$, $x_2$,
and $x_3$, and respective diffusivities $D_i$. It is again
useful to introduce the scaled coordinates $y_i=x_i/\sqrt{D_i}$ which renders
the diffusion in the $y_i$ coordinates spatially isotropic. In terms of
$y_i$, lamb survival corresponds to $y_2\sqrt{D_2}<y_3\sqrt{D_3}$ and
$y_1\sqrt{D_1}<y_3\sqrt{D_3}$. These constraints mean that the effective
particle in three-space remains behind the plane
$y_2\sqrt{D_2}=y_3\sqrt{D_3}$ and to the left of the plane
$y_1\sqrt{D_1}=y_3\sqrt{D_3}$ (Fig.~\ref{wedge}(a)); this geometry
is a wedge region of opening angle $\Theta$ defined by the
intersection of these two planes. If the particle hits one of the
planes, then one of the lions has killed the lamb.

This mapping therefore provides the lamb survival probability, since it is
known that the survival probability of a diffusing particle within this
absorbing wedge asymptotically decays as\cite{cj}
\begin{equation}
\label{wedge-def}
S_{\rm wedge}(t)\sim t^{-\pi/2\Theta}.
\end{equation}
For completeness, we derive this asymptotic behavior
by mapping the diffusive system onto a corresponding electrostatic system in
Appendix~A. To determine the value of $\Theta$ which corresponds to
our 3-particle system, notice that the unit normals to the planes
$y_1\sqrt{D_1}=y_3\sqrt{D_3}$ and $y_2\sqrt{D_2}=y_3\sqrt{D_3}$ are $\hat
{\bf n}_{13}= (-\sqrt{D_1},0,\sqrt{D_3})/\sqrt{D_1+D_3}$ and $\hat {\bf
 n}_{23} = (0,-\sqrt{D_2},\sqrt{D_3})/\sqrt{D_2+D_3}$, respectively.
Consequently $\cos\phi=\hat {\bf n}_{13}\cdot \hat {\bf n}_{23}$
(Fig.~\ref{wedge}(b)), and the wedge angle is $\Theta= \pi - \phi =
\pi-\cos^{-1}[D_3/\sqrt{(D_1+D_3)(D_2+D_3)}]$. If we take $D_1=D_2= D_L$ for
identical lions, and $D_3= D_\ell$, the survival exponent for the lamb is
\begin{equation}
\label{beta2}
\b_2(\CR)={\pi\over{2\Theta}}=\left[2-{2\over\pi}\cos^{-1}{\CR\over{1+\CR}}\right]^{-1},
\end{equation}
where $\CR= D_\ell/D_L$.

\begin{figure}
\epsfxsize=60mm\epsfysize=60mm\centerline{\epsfbox{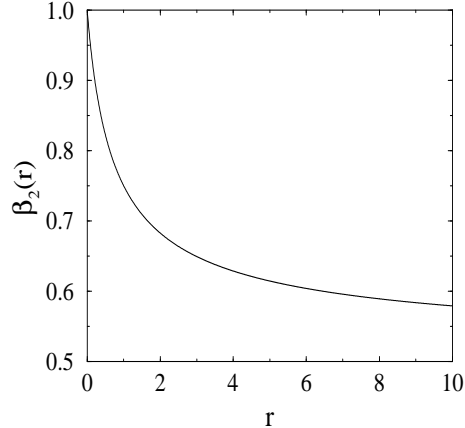}}
\vskip 0.0in
\caption{The survival exponent $\beta_2(r)$ given by
Eq.~(\ref{beta2}) versus the diffusivity ratio $r$.
\label{beta2d}}
\end{figure}

The dependence of $\b_2(r)$ on the diffusivity ratio $\CR$ is shown in
Fig.~\ref{beta2d}. This exponent monotonically decreases from 1 at $\CR=0$ to
1/2 for $\CR\to\infty$. The former case corresponds to a stationary lamb,
where the two lions are statistically independent and $S_2(t)=S_1(t)^2$. On
the other hand, when $\CR\to\infty$ the lamb diffuses rapidly and the motion
of the lions becomes irrelevant. This limit therefore reduces to the
diffusion of a lamb and a stationary absorber, for which $S_2(t)=S_1(t)$.
Finally, for $D_\ell=D_L$, $\b_2=3/4<2\b_1$, and equivalently,
$S_2(t)>S_1(t)^2$. This inequality reflects the fact that the
incremental threat to the lamb from the second lion is less than the
first.

\medskip \noindent {\bf IV. MANY LIONS}

The above construction can, in principle, be extended by
recasting the survival of a lamb in the presence of $N$ lions as
the survival of a diffusing particle in $N+1$ dimensions within an
absorbing hyper-wedge defined by the intersection of the $N$
half-spaces $x_i<x_{N+1}$,
$i=1,2,\ldots,N$. This approach has not led to a tractable
analytical solution. On the other hand, numerical
simulations\cite{bg} indicate that the exponent $\beta_{N}$ 
grows slowly with $N$, with $\b_3\approx 0.91$,
$\b_4\approx 1.03$, and $\b_{10}\approx 1.4$. The understanding of the slow
dependence of $\b_N$ on $N$ is the focus of this section.

\smallskip \noindent{\bf A. LOCATION OF THE LAST LION}

\begin{figure}
\epsfxsize=70mm\epsfysize=70mm\centerline{\epsfbox{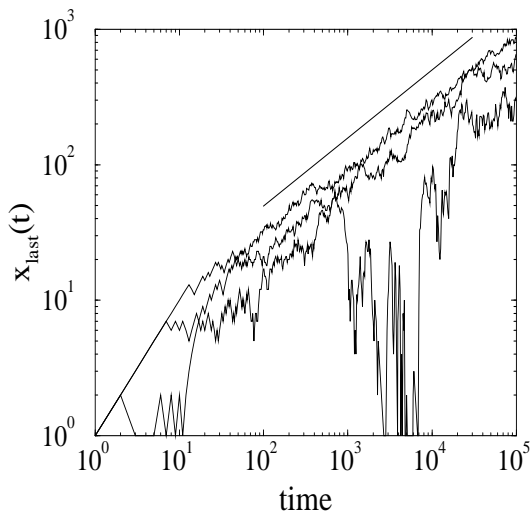}}
\vskip 0.1in
\caption{Time dependence of $x_{\rm last}$ for a single realization of $N=4$, 
64, and 1028 lions (bottom to top). This data was generated by
tracking the position of the rightmost among $N$ lions, each of
which performs a nearest-neighbor discrete-time random walk
starting from $x=0$. The coincidence of the data and the linear
early-time growth of $x_{\rm lst}$ are artifacts of the discrete
random walk motion. The straight line of slope 1/2 indicates the
expected long-time behavior. 
\label{xlast}}
\end{figure}

One way to understand the behavior of the survival probability is to focus on
the lion closest to the lamb, because this last lion ultimately kills the
lamb. As was shown in Fig.~\ref{space-time}, the individual identity of this
last lion can change with time due to the crossing of different lion
trajectories. In particular, crossings between the last lion and its left
neighbor lead to a systematic rightward bias of the last lion. This bias is
stronger for increasing $N$, due to the larger number of crossings of the
last lion, and this high crossing rate also leads to $x_{\rm last}(t)$
becoming smoother as $N$ increases (Fig.~\ref{xlast}). This approach of the
last lion to the lamb is the mechanism which leads to the survival
probability of the lamb decaying as $t^{-\b_N}$, with $\b_N$ a slowly
increasing function of $N$.

To determine the properties of this last lion, suppose that $N\gg
1$ lions are initially at the origin. If the lions perform
nearest-neighbor, discrete-time random walks, then at short times,
$x_{\rm last}(t)=t$. This trivial dependence persists as long as
the number of lions at the last site in their spatial distribution
is much greater than one. In this case there is a large probability
that one of these lions will hop to the right, thus maintaining the
deterministic growth of
$x_{\rm last}$. This growth will continue as long as
${N\over\sqrt{4\pi D_Lt}} e^{-t^2/4D_Lt}\gg 1$, that is, for $t\ll
4D_L\ln N$. At long times, an estimate for the location of the last
lion is provided by the condition\cite{extreme}
\begin{equation}
\label{xlast-defn}
\int_{x_{\rm last}}^\infty {N\over\sqrt{4\pi D_Lt}}\,e^{-x^2/4D_Lt}\,\,dx= 1.
\end{equation}
Equation~(\ref{xlast-defn}) specifies that there is one lion out of
an initial group of
$N$ lions which is in the range $[x_{\rm last},\infty]$. Although
the integral in Eq.~(\ref{xlast-defn}) can be expressed in terms of
the complementary error function, it is instructive to evaluate it
explicitly by writing
$x=x_{\rm last}+\epsilon$ and re-expressing the integrand in terms
of $\epsilon$. We find that
\begin{equation}
\label{xlast-app}
\int_{x_{\rm last}}^\infty {N\over\sqrt{4\pi D_Lt}}\,e^{-x_{\rm last}^2/4D_Lt}\,
e^{-x_{\rm last}\epsilon/2D_Lt}\, e^{-\epsilon^2/4D_Lt}
\,\,d\epsilon= 1.
\end{equation}
Over the range of $\epsilon$ for which the second exponential
factor is non-negligible, the third exponential factor is nearly
equal to unity. The integral in Eq.~(\ref{xlast-app}) thus reduces
to an elementary form, with the result
\begin{equation}
\label{xlast-result}
{N\over\sqrt{4\pi D_Lt}}\,e^{-x_{\rm last}^2/4D_Lt}\,{2D_Lt\over x_{\rm last}}= 1.
\end{equation}

If we define $y=x_{\rm last}/\sqrt{4D_Lt}$ and $M=N/\sqrt{4\pi}$,
the condition in Eq.~(\ref{xlast-result}) can be simplified to 
\begin{equation}
\label{xlast-simp}
ye^{y^2}=M,
\end{equation}
with the solution
\begin{equation}
\label{xlast-final}
y=\sqrt{\ln M\,}\Bigl(1-{1\over 4}{\ln (\ln M) \over{\ln
M}}+\ldots\Bigr).
\end{equation}
In addition to obtaining the mean location of the last lion, extreme
statistics can be used to find the spatial probability of the last lion.  For
completeness, this calculation is presented in Appendix~B.

To lowest order, Eq.~(\ref{xlast-final}) gives
\begin{equation}
\label{xl-N}
x_{\rm last}(t)\sim\sqrt{4D_L\ln N\,t}\equiv\sqrt{A_N t},
\end{equation}
for finite $N$. For $N=\infty$, $x_{\rm last}(t)$ would always equal $t$ if
an infinite number of discrete random walk lions were initially at the
origin. A more suitable initial condition therefore is a 
concentration $c_0$ of lions uniformly distributed from $-\infty$ to
0. In this case, only $N\propto\sqrt{c^2_0 D_L t}$ of the lions are
``dangerous,'' that is, within a diffusion distance from the edge
of the pack and thus potential candidates for killing the lamb.
Consequently, for $N\to\infty$, the leading behavior of $x_{\rm
last}(t)$ becomes
\begin{equation}
\label{xl-inf}
x_{\rm last}(t) \sim \sqrt{2D_L\ln(c^2_0 D_L t)\,t}\, .
\end{equation}
As we discuss in Section~IVB, the survival probability of the lamb
in the presence of many lions is essentially determined by this
behavior of $x_{\rm last}$.

\smallskip \noindent{\bf B. LAMB SURVIVAL PROBABILITY FOR LARGE
$N$}

An important feature of the time dependence of $x_{\rm last}$ is that
fluctuations decrease for large $N$ (Fig.~\ref{xlast}). Therefore the lamb
and $N$ diffusing lions can be recast as a two-body system of a lamb and an
absorbing boundary which {\em deterministically} advances toward the lamb as
$x_{\rm last}(t)=\sqrt{A_Nt}$.

To solve this two-body problem, it is convenient to change coordinates
from $[x,t]$ to $[x'=x-x_{\rm last}(t),t]$ to fix the absorbing boundary at
the origin. By this construction, the diffusion equation for the lamb
probability distribution is transformed to the convection-diffusion equation
\begin{equation}
\label{CDE}
\pd{p(x',t)}{t}- {x_{\rm last}\over 2t}\pd{p(x',t)}{x'}= D\pnd{p(x',t)}{x'}{2}, 
\quad (0\le x' <\infty)
\end{equation}
with the absorbing boundary condition $p(x'=0,t)=0$. In this reference frame
which is fixed on the average position of the last lion, the second term in
Eq.~(\ref{CDE}) accounts for the bias of the lamb towards the absorber with a
``velocity'' $-x_{\rm last}/2t$. Because $x_{\rm last}\sim \sqrt{A_Nt}$ and
$x'\sim\sqrt{D_\ell t}$) have the same time dependence, the lamb survival
probability acquires a nontrivial dependence on the dimensionless parameter
$A_N/D_\ell$.\cite{movingm,movingp,movingkr} Such a dependence is in contrast
to the cases $x_{\rm last}\ll x'$ or $x_{\rm last}\gg x'$, where the
asymptotic time dependence of the lamb survival is controlled by the faster
of these two co-ordinates. Such a phenomenon occurs whenever there is a
coincidence of fundamental length scales in the system (see, \eg,
Ref.~\ref{bib:baren}).

Equation~(\ref{CDE}) can be transformed into the parabolic cylinder equation
by the following steps.\cite{kr} First introduce the dimensionless length
$\xi=x'/x_{\rm last}$ and make the following scaling ansatz for the lamb
probability density,
\begin{equation}
\label{cxt}
p(x',t) \sim t^{-\b_N-1/2}\CC(\xi).
\end{equation}
The power law prefactor in Eq.~(\ref{cxt}) ensures that the integral of
$p(x',t)$ over all space, namely the survival probability, decays as
$t^{-\b_N}$, and $\CC(\xi)$ expresses the spatial dependence of the lamb
probability distribution in scaled length units. This ansatz codifies the
fact that the probability density is not a function of $x'$ and $t$
separately, but is a function only of the dimensionless ratio $x'/x_{\rm
 last}(t)$. The scaling ansatz provides a simple but powerful approach for
reducing the complexity of a wide class of systems with a divergent
characteristic length scale as $t\to\infty$.\cite{baren}

If we substitute Eq.~(\ref{cxt}) into Eq.~(\ref{CDE}), we obtain
\begin{equation}
\label{de-scaled}
{D_\ell\over A_N}\t2d{\CC}{\xi}+{1\over 2}(\xi+1)\td\CC \xi+
\bigl(\b_N+{1\over 2}\bigr)\CC=0.
\end{equation}
Now introduce $\eta=(\xi+1)\sqrt{A_N/2D_\ell}$ and
$\CC(\xi)=e^{-\eta^2/4}\,\CD(\eta)$ in Eq.~(\ref{de-scaled}). This
substitution leads to the parabolic cylinder equation of order
$2\b_N$\cite{bo}
\begin{equation}
\label{pce}
{d^2\CD_{2\b_N}\over d\eta^2}+
\biggl[2\b_N+{1\over 2}-{\eta^2\over 4}\biggr]\CD_{2\b_N}=0,
\end{equation}
subject to the boundary condition, $\CD_{2\b_N}(\eta)=0$ for both
$\eta=\sqrt{A_N/2D_\ell}$ and $\eta=\infty$. Equation~(\ref{pce}) has the form
of a Schr\"odinger equation for a quantum particle of energy $2\beta_N+{1\over
 2}$ in a harmonic oscillator potential $\eta^2/4$ for
$\eta>\sqrt{A_N/2D_\ell}$, but with an infinite barrier at
$\eta=\sqrt{A_N/D_\ell}$.\cite{sho} For the long time behavior, we want the
ground state energy in this potential. For $N\gg 1$, we may approximate this
energy as the potential at the classical turning point, that is,
$2\b_N+{1\over 2}\simeq \eta^2/4$. We therefore obtain $\b_N\sim
A_N/16D_\ell$. Using the value of $A_N$ given in Eqs.~(\ref{xl-N}) and
(\ref{xl-inf}) gives the decay exponent
\begin{equation}
\label{beta-N}
\b_N\sim \cases{{D_L\over{4D_\ell}} \ln N, & \mbox{if}\ $N$ finite\cr\cr
  {D_L\over 8D_\ell} \ln t . & \mbox{for}\ $N=\infty$}
\end{equation}
The latter dependence of $\beta_N$ implies that for $N\to\infty$,
the survival probability has the log-normal form
\begin{equation}
\label{S-inf}
S_{\infty}(t)\sim \exp\left(-{D_L\over{8D_\ell}}\,\ln^2t\right).
\end{equation}

Although we obtained the survival probability exponent $\beta_N$
for arbitrary diffusivity ratio $r=D_\ell/D_L$, simple
considerations give different behavior for $\CR\gg 1$
and $\CR\ll 1$. For example, for $r=0$ (stationary lamb) the
survival probability decays as $t^{-N/2}$. Therefore
Eq.~(\ref{beta-N}) can no longer apply for $\CR< N^{-1}$, where
$\b_N(\CR)$ becomes of order $N$. Conversely, for $r=\infty$
(stationary lions), the survival probability of the lamb decays as
$t^{-1/2}$. Thus Eq.~(\ref{beta-N}) will again cease to be valid
for $\CR>\ln N$, where
$\b_N(\CR)$ becomes of order unity. By accounting for these limits,\cite{kr}
the dependence of $\b_N$ on the diffusivity ratio $\CR$ is expected to be
\begin{equation}
\label{beta-R}
\b_N(\CR)=\cases{N/2 &\qquad $\CR\ll 1/N$ \cr
 (1/4\CR) \ln(4N) &\qquad $1/N\ll \CR\ll \ln N$ \cr
 1/2 &\qquad $\CR\gg \ln N$. \cr}
\end{equation}
The $r$ dependence of $\b_N$ in the intermediate regime of $1/N\ll
\CR\ll \ln N$ generalizes the exponents given in Eq.~(\ref{beta2})
for the three-particle system to general $N$.

\medskip \noindent{\bf V. DISCUSSION}

We investigated diffusive capture of a lamb in one dimension by
using of several essential techniques of nonequilibrium statistical
mechanics including first-passage properties of diffusion, extreme
value statistics, electrostatic analogies, scaling analysis, and
moving boundary value theory. These tools provide an appealing
physical description for the survival probability of the lamb in
the presence of $N$ lions for the cases of very small and very large
$N$. Nevertheless, the exact solution to capture of the lamb
remains elusive for $N\geq 3$.

We close by suggesting several avenues for further study:
\begin{enumerate}
 
\item {\em Better Simulation Methods.} Previous simulations of this
 system\cite{bg} followed the random walk motion of one lamb and $N$ lions
 until the lamb was killed. This type of simulation is simple to construct.
 One merely places the lamb and lions on a one-dimensional lattice and have
 them perform independent nearest-neighbor random walks until one lion lands
 on the same site as the lamb. The survival probability is obtained by
 averaging over a large number of realizations of this process. However,
 following the motion of discrete random walks is inefficient, because it is
 unlikely for the lamb to survive to long times and many realizations of the
 process must be simulated to obtain accurate long-time data. In principle,
 a much better approach would be to propagate the exact probability
 distribution of the particles in the system.\cite{hav} Can such an approach
 be developed for the lamb-lion capture process? Another possibility is to
 devise a simple discrete random-walk process to simulate the motion of the
 last lion. Such a construction would permit consideration of just the lamb
 and the last lion, thus providing significant computational efficiency.

\item {\em The Last Lion.} Extreme value statistics provides the spatial
  probability distribution of the last lion. We may also ask other basic
  questions: How long is a given lion the ``last'' one? How many lead changes
  of the last lion occur up to time $t$?  How many different lions may be in
  the lead up to time $t$? What is the probability that a particular lion is
  never in the lead?  Methods to investigate some of these issues are also
  outlined in the article by Schmittmann and Zia\cite{sz} in this journal
  issue.

\item {\em Spatial Probability Distribution of the Lamb.} As we have seen for
 the case of one lion, the spatial distribution of the lamb is a useful
 characteristic of the capture process. What happens for large $N$? In
 principle, this information is contained in the solution to the parabolic
 cylinder equation for the scaled probability distribution
 (Eq.~(\ref{pce})). The most interesting behavior is the form of the
 distribution close to the absorbing boundary, where the interaction between
 the lions and the lamb is strongest. For $N=1$ lion, this distribution
 decays linearly to zero as a function of the distance to the lion, while
 for $N=2$, the distribution has a power law decay in the distance to the
 last lion which depends on the diffusivity ratio $D_\ell/D_L$. What happens
 for general $N$ and for general diffusivity ratio? Is there a physical way
 to determine this behavior?
 
\item {\em Two-Sided Problem.} If $N$ lions are located on both sides of the
 lamb, then the lamb is relatively short-lived because there is no escape
 route. One can again construct a mapping between the $N+1$-particle
 reacting system and the diffusion of an effective particle in an absorbing
 wedge-shaped domain in $N+1$ dimensions. From this mapping, $S_N(t)$
 decays as $t^{-\gamma_N}$, but the dependence of $\gamma_N$ on $N$ and
 diffusivity ratio is unknown. It is clear, however, that the optimal
 strategy for the surrounded lamb is to remain still, in which case the
 lions are statistically independent and we then recover $S_N(t)\sim
 t^{-N/2}$. Is there a simple approach that provides the dependence of
 $\gamma_N$ on $N$ for arbitrary diffusivity ratio? Finally, $S_\infty(t)$
 exhibits a stretched exponential decay in time,
 $\exp(-t^{1/2})$.\cite{rk,bzk} What is that nature of the transition from
 finite $N$ to infinite $N$ behavior?
 
\item {\em Intelligent Predators and Prey.} In a more realistic capture
 process, lions would chase the lamb, while the lamb would attempt to run
 away. What are physically-reasonable and analytically-tractable rules for
 such directed motion which would lead to new and interesting kinetic
 behaviors?

\end{enumerate}
\medskip \noindent{\bf ACKNOWLEDGMENTS}

We gratefully acknowledge NSF grant DMR9632059 and ARO grant
DAAH04-96-1-0114 for partial support of this research.

\medskip \noindent{\bf APPENDIX A:  SURVIVAL PROBABILITY OF A
DIFFUSING PARTICLE WITHIN A WEDGE}

The survival probability of a diffusing particle within an absorbing wedge
can be derived by solving the diffusion equation in this
geometry.\cite{dba,cj} We provide an alternative derivation of this result
can also be obtained by developing a correspondence between diffusion and
electrostatics in the same boundary geometry. Although the logic underlying
the correspondence is subtle, the result is simple and has wide
applicability.

The correspondence rests on the fact that the integral of the diffusion
equation over all time reduces to the Poisson equation. This time integral is
\begin{equation}
\label{de-wedge}
\int_0^\infty \left\{D\nabla^2 p(r,\theta,t)= \pd{p(r,\theta,t)}{t}\right\}dt.
\end{equation}
If one defines an electrostatic potential by
$\Phi(r,\theta)=\!\int_0^\infty p(r,\theta,t)\,dt$,
Eq.~(\ref{de-wedge}) can be written as
\begin{equation}
\label{poisson}
\nabla^2 \Phi(r,\theta)=-{1\over D}\left[p(r,\theta,t=\infty)-p(r,\theta,t=0)\right].
\end{equation}
For a boundary geometry such that the asymptotic survival probability in the
diffusive system is zero, then Eq.~(\ref{poisson}) is just the Poisson
equation, with the initial condition in the diffusive system corresponding to
the charge distribution in the electrostatic system, and with absorbing
boundaries in the diffusive system corresponding to grounded
conductors.

To exploit this analogy, we first note that the electrostatic potential in
the wedge decays as $\Phi(r,\theta)\sim r^{-\pi/\Theta}$ for $r\to\infty$,
for any localized charge distribution.\cite{jackson} Because the
survival probability of a diffusing particle in the absorbing wedge
is given by
$S(t)=\!\int p(r,\theta,t)\, dA$, where the integral is over the
area of the wedge, we find the following basic relation between
$S(t)$ and the electrostatic potential in the same boundary geometry
\begin{eqnarray}
\label{de-la}
\int_0^t S(t)\, dt&=& \! \int_0^t \! dt \! \int \! p(r,\theta,t)\,dA
\nonumber \\
 &\approx& \! \int_0^{\sqrt{Dt}} r\,dr \! \int_0^\Theta \! d\theta\,\,
\Phi(r,\theta) \nonumber \\
 &\sim& \! \int_0^{\sqrt{Dt}} r^{1-\pi/\Theta} \,dr \nonumber \\
 &\propto & t^{1-\pi/2\Theta}.
\end{eqnarray}
In evaluating the time integral of the survival probability, we use the fact
that particles have time to diffuse to radial distance $\sqrt{Dt}$ but no
further. Thus in the second line of Eq.~(\ref{de-la}), the time integral of
the probability distribution reduces to the electrostatic potential for
$r<\sqrt{Dt}$ and is essentially zero for $r>\sqrt{Dt}$. Finally, by
differentiating the last equality in Eq.~(\ref{de-la}) with respect to time
we recover Eq.~(\ref{wedge-def}).

\medskip \noindent{\bf APPENDIX B:  SPATIAL DISTRIBUTION OF THE
LAST LION BY EXTREME STATISTICS}

It is instructive to apply extreme statistics to determine the probability
distribution for the location of the last lion from an ensemble of $N$
lions.\cite{extreme} Let $p(x)={1\over\sqrt{4\pi D_Lt}} e^{-x^2/4D_Lt}$ be
the (Gaussian) probability distribution of a single lion. Then
$p_>(x)\equiv\!\int_x^\infty p(x')\,dx'$ is the probability that a
diffusing lion is in the range $[x,\infty]$ and similarly
$p_<(x)=1-p_>(x)$ is the probability that the lion is in the range
$[-\infty,x]$. Let $L_N(x)$ be the probability that the last lion
out of a group of $N$ is located at $x$. This extremal probability
is given by
\begin{equation}
\label{LN}
L_N(x) = Np(x)p_<(x)^{N-1}.
\end{equation}
That is, one of the $N$ lions is at $x$, while the remaining $N-1$ lions are
in the range $[-\infty,x]$. If we evaluate the factors in
Eq.~(\ref{LN}), we obtain a double exponential
distribution:\cite{extreme}
\begin{eqnarray}
\label{LN-exp}
L_N(x) &=& {N\over\sqrt{4\pi D_Lt}}e^{-x^2/4D_Lt}
\left[1-\!\int_x^\infty \!{1\over\sqrt{4\pi D_Lt}}
e^{-x^2/4D_Lt} \,dx\right]^{N-1}, \nonumber \\
 &\sim& {N\over\sqrt{4\pi D_Lt}}e^{-x^2/4D_Lt} 
\exp\left[- {N-1\over\sqrt{4\pi D_Lt}} \!\int_x^\infty \!
e^{-x^2/4D_Lt}\,dx\right].
\end{eqnarray}
When $N$ is large, then $x/\sqrt{4D_Lt}$ is also large, and we can
asymptotically evaluate the integral in the exponential in
Eq.~(\ref{LN-exp}). Following Eq.~(\ref{xlast-simp}), it is
convenient to express the probability distribution in terms of $M=
N/\sqrt{4\pi}$ and
$y=x_{\rm last}/\sqrt{4D_Lt}$. If we use $L_N(y)\,dy=L_N(x)\,dx$, we
obtain
\begin{equation}
\label{LN-final}
L_N(y)\simeq 2M e^{-y^2}\exp(-Me^{-y^2}/y).
\end{equation}
The most probable value of $x_{\rm last}$ is determined by the requirement
that $L_N'(y)=0$. This condition reproduces $ye^{y^2}=M$ given in
Eq.~(\ref{xlast-simp}).

We may also estimate the width of the distribution from its inflection
points, that is, when $L_M''(y)=0$. By straightforward calculation,
$L_N''(y)=0$ at
\begin{equation}
\label{ypm}
y_\pm\simeq\sqrt{\ln(M/k_\pm)}\approx\sqrt{\ln M}
\Bigl(1-{\ln k_\pm\over{\ln M}}\Bigr),
\end{equation}
where $k_\pm=(3\pm\sqrt{5})/2$. Therefore as $N\to\infty$, the width of
$L_N(y)$ vanishes as $1/\sqrt{\ln M}$. This behavior is
qualitatively illustrated in Fig.~\ref{xlast}, where the
fluctuations in $x_{\rm last}(t)$ decrease dramatically as $N$
increases. This decrease can also be understood from the form of
the extreme distribution
$L_N(x)$ in Eq.~(\ref{LN-exp}). The large-$x$ decay of $L_N(x)$ is
governed by $p(x)$, while the double exponential factor becomes an
increasingly sharp step at $x_{\rm step}\approx\sqrt{4D_Lt\ln N}$
as $N$ increases. The product of these two factors leads to
$L_N(x)$ essentially coinciding with $p(x)$ for $x>x_{\rm step}$
and $L_N(x)\approx 0$ for $x<x_{\rm step}$.


\begin{thebibliography}{99}

\bibitem{bg} M.~Bramson and D.~Griffeath, ``Capture problems for coupled
 random walks,'' in {\sl Random walks, Brownian Motion, and Interacting
 Particle Systems: A Festschrift in Honor of Frank Spitzer}, 
 R.~Durrett and H.~Kesten, eds. (Boston, Birkhauser, 1991), pp.\
153--188.

\bibitem{kesten} H. Kesten, ``An Absorption Problem for Several Brownian
 Motions,'' in {\sl Seminar on Stochastic Processes, 1991}, E. \c
 Cinlar, K. L. Chung, and M. J. Sharpe, eds.\ (Birkh\"auser,
Boston, 1992).
 
\bibitem{kr} P. L. Krapivsky and S. Redner, ``Kinetics of a diffusive
 capture process: lamb besieged by a pride of lions,'' \jpa 29 5347--5357
 1996.
 
\bibitem{feller} W.~Feller, {\sl An Introduction to Probability
Theory} (J. S. Wiley \& Sons, New York, 1971), Vol.~1.

\bibitem{weiss} G. H. Weiss, {\sl Aspects and Applications of the Random Walk}
 (North-Holland, Amsterdam, 1994).~\label{bib:weiss}
 
\bibitem{dba} D.~ben-Avraham, ``Computer simulation methods for
 diffusion-controlled reactions,'' {\sl J.\ Chem.\ Phys.} {\bf 88}, 941--947
 (1988); M.~E.~Fisher and M.~P.~Gelfand, ``The reunions of three dissimilar
 vicious walkers,'' {\sl J. Stat.\ Phys.} {\bf 53}, 175--189
(1988).~\label{bib:dba}
 
\bibitem{extreme} For a general introduction to the statistics of
extremes, see J. Galambos, {\sl The Asymptotic Theory of Extreme
Order Statistics} (R. E. Krieger Pub., Malabar, FL, 1987). A more
recent discussion of extreme statistics with applications to many
types of records, see B. C. Arnold, N. Balakrishnan, and H. N.
Nagaraja, {\sl Records} (J. S. Wiley \& Sons, New York, 1998).

\bibitem{movingm} L.~Breiman, ``First exit time from the square root
 boundary,'' {\sl Proc.\ Fifth Berkeley Symp.\ Math.\ Statist.\ and Probab.}
 {\bf 2}, 9--16 (1966); H.~E.~Daniels, ``The minimum of a stationary Markov
 superimposed on a U-shape trend,'' {\sl J.\ Appl.\ Prob.} {\bf 6}, 399--408
 (1969); K.~Uchiyama, ``Brownian first exit from sojourn over one sided
 moving boundary and application,'' {\sl Z. Wahrsch.\ verw.\ Gebiete} {\bf
 54}, 75--116 (1980); P.~Salminen, ``On the hitting time and the exit time
 for a Brownian motion to/from a moving boundary,'' {\sl Adv.\ Appl.\ Prob.}
 {\bf 20}, 411--426 (1988).
 
\bibitem{movingp} L.~Turban, ``Anisotropic critical phenomena in parabolic
 geometries: The directed self-avoiding walk,'' {\sl J. Phys.\ A} {\bf 25},
 L127--L134 (1992); F.~Igl\'oi, ``Directed polymer inside a parabola: Exact
 solution,'' {\sl Phys.\ Rev.\ A} {\bf 45}, 7024--7029 (1992); F.~Igl\'oi,
 I.~Peschel, and L.~Turban, ``Inhomogeneous systems with unusual critical
 behaviour,'' {\sl Adv.\ Phys.} {\bf 42}, 683--740 (1993).
 
\bibitem{movingkr} P.~L.~Krapivsky and S.~Redner, ``Life and death in an
 expanding cage and at the edge of a receding cliff,'' \ajp 64 546--552
1996.
 
\bibitem{as} M. Abramowitz and I. A. Stegun, {\sl Handbook of
Mathematical Functions} (U.S. Govt.\ Printing Office, Washington,
1972).
 
\bibitem{cj} H. S. Carslaw and J. C. Jaeger, {\sl Conduction of Heat in
 Solids} (Oxford University Press, Oxford, 1959), Chap.\ XI.
 
\bibitem{baren} G. I. Barenblatt, {\sl Scaling, Self-similarity,
and Intermediate Asymptotics} (Cambridge University Press,
Cambridge, 1996).~\label{bib:baren}
 
\bibitem{bo} C.~M.~Bender and S.~A.~Orszag, {\sl Advanced Mathematical
 Methods for Scientists and Engineers} (McGraw-Hill, New York, 1978).

\bibitem{sho} See, for example, L. I. Schiff, {\sl Quantum Mechanics},
 (McGraw-Hill, New York, 1968).
 
\bibitem{hav} For an introduction to this simulation approach, see \eg, S.
 Havlin and D. ben-Avraham, ``Diffusion in Disordered Media,'' \ap 36
 695--798 1987.

\bibitem{sz} B. Schmittmann and R. Zia, ``Weather Records: Musings on Cold
Days and a Long Hot Indian Summer'' cond-mat/9905103.

\bibitem{rk} S.~Redner and K.~Kang, ``Kinetics of the scavenger reaction,''
 {\sl J. Phys.\ A} {\bf 17}, L451--L455 (1984).

\bibitem{bzk} A.~Blumen, G.~Zumofen, and J.~Klafter, ``Target Annihilation by
 Random Walkers,'' {\sl Phys.\ Rev.\ B} {\bf 30}, 5379--5382 (1984).

\bibitem{jackson} J. D. Jackson, {\sl Classical Electrodynamics} (J. S. Wiley
 \& Sons, Inc., New York, 1999), 3rd edition.
 
\end{thebibliography}
\end{document}